\documentclass[11pt,preprintnumbers,titlepage,nofootinbib]{revtex4-2}%
\usepackage[latin9]{inputenc}
\usepackage{amsmath}
\usepackage{amssymb}
\usepackage{dcolumn}
\usepackage{graphicx}
\usepackage[colorlinks=true, pdfstartview=FitV, linkcolor=blue, citecolor=red, urlcolor=magenta]{hyperref}
\usepackage{amsfonts}
\usepackage{subfigure}
\usepackage{cleveref}
\usepackage{listings}
\usepackage{xcolor}
\usepackage{array}
\usepackage{url}
\usepackage{color}%
\usepackage{tikz}
\usepackage{pgfplots}
\usepackage{todonotes}
\usepackage{multirow}

\usepackage{booktabs} 

\usepackage{xcolor}
\usepackage{soul}

\graphicspath{{Figures/}}

\begin{document}

\newcommand{\naveencroatia}{\affiliation { Division of Theoretical Physics,  Rudjer Bo\v skovi\'c Institute,\\ Bijeni\v cka c.54, 
HR-10002 Zagreb, Croatia}}

\newcommand{\nitk}{\affiliation {Department of Physics, National Institute of Technology Karnataka (NITK) Surathkal, Mangaluru - 575025, India}}

\newcommand{\iitropar}{\affiliation {Department of Physics, Indian Institute of Technology, Ropar, Rupnagar, Punjab 140 001, India}}

\newcommand{\shreyas}{\affiliation {Department of Oral Health Sciences, School of Dentistry,\\ University of Washington, Seattle, WA 98195, USA.}}
\title{Thermodynamics, Phase Transition and Joule Thomson Expansion  of 4-D Gauss-Bonnet AdS Black Hole}

\author{Kartheek Hegde}
\email{hegde.kartheek@gmail.com}
\nitk
\author{A. Naveena Kumara}
\email{naviphysics@gmail.com}
\nitk
\naveencroatia
\author{Ahmed Rizwan C.L.}
\email{ahmedrizwancl@gmail.com}
\nitk
\author{Md Sabir Ali}
\email{alimd.sabir3@gmail.com}
\iitropar
\author{Ajith K.M.}
\email{ajithkm@gmail.com}
\nitk
\author{Shreyas Punacha}
\email{shreyas4@uw.edu}
\shreyas
\date{\today}

\begin{abstract}
We explore the thermodynamic and phase transition properties of asymptotically AdS black holes within Einstein-Gauss-Bonnet gravity, focusing on Joule Thomson expansion. Thermodynamics is studied in the extended phase space, where the cosmological constant serves as thermodynamic pressure. We observe that the black hole undergoes a phase transition similar to that of a van der Waals system. We analyze charged and neutral cases separately to distinguish the effect of charge and Gauss Bonnet parameter on critical behavior and examine the phase structure. We find that the Gauss-Bonnet coupling parameter behaves similarly to black hole charge or spin, guiding the phase structure. To understand the underlying phase structure determined by the Gauss-Bonnet coefficient $\alpha$, we introduce a new order parameter. We discover that the change in the conjugate variable to the Gauss-Bonnet parameter acts as an order parameter, demonstrating a critical exponent of $1/2$ in the vicinity of the critical point. Since the phase structure is analogous to that of a van der Waals fluid, we investigate the Joule-Thomson expansion of the black hole. We analytically study the Joule-Thomson expansion, focusing on three key characteristics: the Joule-Thomson coefficient, inversion curves, and isenthalpic curves. We obtain isenthalpic curves in the $T-P$ plane and illustrate the cooling-heating regions.

\end{abstract}
\maketitle





\section{Introduction}
The understanding that black holes behave as thermodynamic systems has been established for several decades \cite{Hawking:1974sw, Bekenstein1972, Bekenstein1973, Bardeen1973}. Notably, the thermodynamics of Anti-de Sitter (AdS) black holes has garnered significant attention owing to the AdS/CFT correspondence \cite{Maldacena:1997re, Gubser:1998bc, Witten:1998qj}. The exploration of the thermodynamic properties of AdS black holes started with the seminal work of Hawking and Page \cite{Hawking:1982dh}, who identified a phase transition in the phase space of the (non-rotating, uncharged) Schwarzschild-AdS black hole. Due to their thermal stability, AdS black holes exhibit thermodynamic characteristics distinct from those of asymptotically flat or de Sitter black holes. Consequently, investigations into phase transitions and critical phenomena have extended to a variety of more complex AdS backgrounds. A significant development was the discovery of a first-order phase transition in the charged (non-rotating) Reissner-Nordstrom AdS (RN-AdS) black hole spacetime \cite{Chamblin:1999tk, Chamblin:1999hg}, exhibiting classical critical behavior reminiscent of a liquid-gas phase transition. Moreover, the identification of the cosmological constant as the thermodynamic variable pressure \cite{Kastor:2009wy, Dolan:2011xt} has initiated the field of black hole chemistry, finding diverse applications in understanding black hole phase transition properties \cite{Kubiznak2012, Gunasekaran2012, Kubiznak:2016qmn} (for comprehensive details, see the review article \cite{Kubiznak:2016qmn} and references therein). An important outcome from the black hole chemistry perspective is the recognition that AdS black holes undergo phase transitions analogous to those in conventional thermodynamic systems. Particularly, AdS black holes exhibit van der Waals (vdW)-like phase transitions \cite{Chamblin:1999tk, Kubiznak2012}, reentrant phase transitions \cite{Altamirano:2013ane}, isolated critical points \cite{Dolan:2014vba, Ahmed:2022kyv}, superfluid-like behavior \cite{Kubiznak:2016qmn}, and multicritical points \cite{Tavakoli:2022kmo}. Black holes displaying van der Waals-like phase structures exhibit the Joule-Thomson expansion, a well-known feature of van der Waals fluids in conventional thermodynamics. Initially, the Joule-Thomson expansion was investigated for charged AdS black holes \cite{Okcu:2016tgt,Okcu:2017qgo, AhmedRizwan:2019yxk}, and it is widely recognized that any AdS black hole with a van der Waals-like phase structure undergoes this process.

In recent years, there has been a growing interest in studying $4D$ Gauss-Bonnet gravity theories (see review article \cite{Fernandes:2022zrq}). Incorporating Einstein-Hilbert and Gauss-Bonnet terms in the gravitational action leads to theories known as Einstein-Gauss-Bonnet gravity. Such theories are intriguing because string theory predicts that classical Einstein's equations are subject to next-to-leading-order corrections, typically described by higher-order curvature terms in the action \cite{Fernandes:2022zrq}. Gauss-Bonnet gravity extends general relativity (GR) by supplementing the Einstein-Hilbert action with a curvature-squared term, 
\begin{equation}
\label{lgb}
\mathcal{G} = R^2 - 4R_{\mu \nu }R^{\mu \nu} + R_{\mu \nu \rho \sigma} R^{\mu \nu \rho \sigma}.
\end{equation}
In $D = 4$, the Gauss-Bonnet term in Eq. \ref{lgb} is purely topological, rendering the full theory indistinguishable from GR. Consequently, the Gauss-Bonnet term in $D = 4$ has long been believed to have no impact on the phenomenology of the classical theory. This belief stems from the Chern theorem, which states that the contribution of $\mathcal{G}$ to the action obtained by integrating the Lagrangian $\mathcal{L}=-2\Lambda+R+\alpha \mathcal{G}$, where $\alpha$ is the coupling constant of the Gauss-Bonnet term, is entirely equivalent to a constant, proportional to the Euler characteristic of the space-time manifold \cite{chern1945curvatura}. Thus, it was presumed to make no contribution to the field equations of the theory, a conclusion that can be explicitly verified using dimensionally dependent identities for the curvature tensors \cite{Edgar:2001vv}.

Recently, considerable effort has been devoted to reformulating Gauss-Bonnet gravity in four dimensions, similar to the efforts made for Einstein gravity in two dimensions \cite{Mann:1992ar}, in order to evoke non-trivial dynamics \cite{Glavan:2019inb, Hennigar:2020lsl, Fernandes:2020nbq, Lu:2020iav}. Such a theory would serve as the only competitor to general relativity (GR) as a metric theory of gravity in four dimensions, featuring equations of motion that are second-order and free of ghost instabilities. In an attempt to introduce the Gauss-Bonnet term directly into four-dimensional gravity, Glavan and Lin proposed rescaling the coupling constant $\alpha$ by $\alpha \rightarrow \alpha/(D-4)$ \cite{Glavan:2019inb}. Although this rescaled quantity becomes divergent as $D \to 4$, Glavan and Lin suggested that by incorporating this rescaling into the Lanczos tensor \cite{lanczos1932elektromagnetismus, lanczos1938remarkable} associated with the Lagrangian $\mathcal{L}$, the terms containing this quantity might remain finite and non-zero. Their postulation was that this divergence could offset the tendency of additional terms in $\mathcal{L}$ to approach zero as $D \rightarrow 4$, potentially allowing the Gauss-Bonnet term to directly influence the four-dimensional theory of gravity. However, the novel approach of Glavan and Lin has faced considerable scrutiny \cite{Gurses:2020ofy, Gurses:2020rxb, Arrechea:2020evj, Arrechea:2020gjw, Bonifacio:2020vbk, Ai:2020peo, Mahapatra:2020rds, Hohmann:2020cor, Cao:2021nng}. Several alternative formulations have emerged, which have garnered significant interest over a short period \cite{Konoplya:2020bxa, Guo:2020zmf, Wei:2020ght, Casalino:2020kbt, Konoplya:2020qqh, Fernandes:2020rpa, Fernandes:2021dsb}. (For further insights and discussions on related works, readers are referred to the review article \cite{Fernandes:2022zrq}). 

There were several efforts to find a consistent Gausss Bonnet theory in 4 dimensions. One of the such efforts was to reconcile 4D Gauss Bonnet solutions from certain scalar-tensor theories, with the likely presence of an additional scalar degree of freedom alongside the two tensor degrees of freedom in the graviton \cite{Bonifacio:2020vbk}. In such theories, it is feasible to work within $D =4$ spacetime dimensions while still incorporating non-vanishing contributions from the Gauss-Bonnet term in the field equations, thanks to the presence of a dilatonic scalar field $\varphi$. It is noteworthy that in terms of purely geometric terms, only couplings of the scalar field to the Ricci scalar and the Gauss-Bonnet term are permissible according to Horndeski's theory. However, in such a scalar-tensor theory, the absence of a quadratic kinetic term for the scalar field results in infinitely strong coupling of the gravitational degrees of freedom \cite{Kobayashi:2020wqy}. A consistent description of the 4D Gauss-Bonnet theory, which differs from general relativity, is provided in \cite{Aoki:2020lig} (See also \cite{Aoki:2020iwm, Aoki_2021}). Using the ADM decomposition, it was demonstrated that the regularization either: (1) breaks the diffeomorphism invariance, resulting in a specific vacuum and implying the absence of a scalar-field degree of freedom, or (2) introduces an additional degree of freedom in the form of a scalar field, consistent with the Lovelock theorem \footnote{Historically, such endeavors can be traced back to the formulation of alternate theories of Einstein gravity. Among the myriad of possibilities, the Einstein-Gauss-Bonnet theory holds a unique position. Initially proposed by Lanczos \cite{lanczos1932elektromagnetismus, lanczos1938remarkable} and subsequently generalized by Lovelock \cite{lovelock1970divergence, lovelock1971einstein}, these theories stand out for requiring no additional fundamental fields beyond those present in GR. Additionally, they maintain the property that the field equations of the theory can be expressed with no higher than second derivatives of the metric - a condition sufficient to avoid Ostrogradsky instability \cite{ostrogradski1850memoiresurlesequationsdifferentiellesrelatives}.}. Following this approach, we adopt a properly defined $4D$ Gauss-Bonnet solution in our article.

It is natural to explore the theoretical and observational implications of this modified classical gravity theory. In this article, we aim to investigate various thermal properties of the $4D$ Gauss-Bonnet black hole in the AdS background, considering both the charged and neutral cases. While the thermodynamics and phase transition properties of higher-dimensional Gauss-Bonnet black holes have been well established for some time, it is not necessarily the case that all phase structure properties generalize to $4D$ Gauss-Bonnet theories. In higher-dimensional Gauss-Bonnet theories, a variety of phase transition properties are observed, including van der Waals (vdW) type phase transitions. However, we shall demonstrate that the $4D$ Gauss-Bonnet black hole exhibits only a vdW-like phase transition. This situation is similar to that of the Kerr-AdS case, where a vdW phase transition is observed in $4D$, while higher dimensions display reentrant and other complex phase transitions. We begin by investigating the thermal properties in the $4D$ Gauss-Bonnet black hole spacetime, considering the charged and neutral cases separately to discern the effect of charge on critical behavior and to examine the phase structure in the absence of charge. Our analysis reveals that the Gauss-Bonnet coupling parameter induces effects similar to those of black hole charge or spin in guiding the black hole phase structure. Consequently, to comprehend the underlying phase structure determined by the Gauss-Bonnet coefficient $\alpha$, we introduce a new order parameter. Given the similarity of the phase structure to that of a vdW fluid, we proceed to investigate the Joule-Thomson expansion of the black hole.

The paper is organized as follows: In the next section, we delve into the elementary thermodynamics of the black hole, starting from the metric details. Section \ref{sectwo} is dedicated to the study of the phase transitions and critical behavior of the black hole. Following that, in section \ref{secthree}, we explore the Joule-Thomson expansion of the black hole system. Finally, in section \ref{secfour}, we conclude the paper with discussions and insights on the results obtained.


\section{Thermodynamics of the Black Hole}
\label{secone}
In this section, we investigate the thermodynamic properties of the $4D$ Gauss-Bonnet AdS black hole, examining both neutral and charged scenarios. As outlined in the introduction, our approach relies on the $4D$ Gauss-Bonnet solutions derived from the regularization method focusing on breaking temporal diffeomorphism \cite{Aoki:2020lig}. The $D$-dimensional covariant action of Einstein-Gauss-Bonnet gravity is given by \cite{Aoki:2020lig},
\begin{equation}
S=\frac{1}{2\kappa _D ^2} \int d^D x \sqrt{-g} \left[ R-\Lambda  +\alpha \mathcal{G}\right] .
\end{equation}
Here, $\alpha$ denotes the Gauss-Bonnet coupling parameter, $\Lambda$ is the cosmological constant, $\mathcal{G}$ is the Gauss-Bonnet invariant (as given in \eqref{lgb}), $\kappa _D$ is the gravitational coupling constant and the dimension $D =d+1$. The spherically symmetric solution is given by,
\begin{equation}
ds^2=-f(r)dt^2+\frac{1}{f(r)}dr^2+r^2d\Omega ^2_{D-2}.
\end{equation}
The solution to the metric function $f(r)$ exhibits two distinct branches: one with a positive sign, referred to as the Gauss-Bonnet branch, and another with a negative sign, known as the GR branch. These branches are well-known in higher-dimensional Einstein-Gauss-Bonnet theory (see, for instance, \cite{Boulware:1985wk}) and persist as a characteristic feature in $4D$ Einstein-Gauss-Bonnet theory. Notably, the positive branch lacks a well-defined limit as $\alpha$ approaches zero, while the negative branch converges to the dynamics of GR. Consequently, the positive branch is labeled the Gauss-Bonnet branch, and the negative one is termed the GR branch. The metric function corresponding to the GR branch is given by \footnote{The charged version of the theory in the AdS background can be obtained by incorporating the Maxwell term $F^{\mu \nu}F_{\mu \nu}$, leading to the Einstein-Maxwell Gauss-Bonnet gravity \cite{Fernandes:2020rpa}.} \footnote{This solution also arises in the context of conformal anomaly gravity \citep{Cai:2009ua, Cai:2014jea}.},
\begin{equation}
f(r)=1+\frac{r^2}{2\alpha} \left(1-\sqrt{1+4 \alpha  \left(-\frac{1}{l^2}+\frac{2 M}{r^3}-\frac{Q^2}{r^4}\right)}\right)
\end{equation}
Here, $M$ and $Q$ denote the mass and charge of the black hole, respectively, while $l$ represents the AdS length, related to the cosmological constant as $\Lambda =-\frac{3}{l^2}$. Both charged and neutral Gauss-Bonnet black holes exhibit multiple inner and outer horizons. The event horizon $r_h$ of the black hole is determined by the largest root of the equation $f(r_h)=0$, representing the norm of the time-like Killing vector associated with the time coordinate $t$. In addition to the horizon radius $r_h$, the Gauss-Bonnet black hole system is characterized by other length scales, such as the AdS radius length $l$, charge $Q$, and the Gauss-Bonnet parameter $\alpha$. Below, the thermodynamic variables of the black hole are expressed in terms of these parameters. The explicit expression for the mass of the black hole is obtained by using the condition $f(r_h)=0$,
\begin{equation}
M=\frac{1}{2}\left(\frac{r_h^3}{ l^2}+\frac{Q^2}{r_h}+\frac{\alpha }{ r_h}+r_h\right).
\end{equation}
We explore the thermodynamics of the black hole within the extended phase space framework, where the cosmological constant is considered a thermodynamic variable-pressure-utilizing the relation $P=-\frac{\Lambda}{8\pi}$. The Hawking temperature of the black hole is associated with the surface gravity $\kappa$, computed through the standard Euclidean trick,
\begin{equation}
T=\frac{\kappa}{2\pi}=\frac{f'(r_h)}{4\pi}=-\frac{\alpha -8 \pi  P r_h^4+Q^2-r_h^2}{4 \pi  r_h^3+8 \pi  \alpha  r_h}.
\label{Hawking}
\end{equation}
It is noteworthy that the Gauss-Bonnet parameter $\alpha$ appears in this expression, indicating the modification of the Hawking temperature in Gauss-Bonnet spacetime. With these thermodynamic variables, we can formulate the first law for the black hole \footnote{We emphasize that all thermodynamic quantities, along with the explicit form of the first law, can be derived by computing the free energy from the on-shell Euclidean action. It is worth noting that both methods yield identical results, thereby serving as a straightforward consistency check for the results presented in this section},
\begin{equation} \label{firstlaw}
dM=TdS+VdP+\Phi dQ+ \mathcal{A}d\alpha 
\end{equation} 
Here, $\Phi$ and $\mathcal{A}$ represent the potentials conjugate to $Q$ and $\alpha$, respectively. From this, we can derive the black hole entropy as, \footnote{During the preparation of this manuscript, we came across a study arguing against the presence of logarithmic correction terms in scalar tensor theories \cite{Liska:2023fdz}. It is important to note, however, that the qualitative thermodynamic behavior remains unchanged. This is because, the different entropy prescriptions are essentially a matter of scaling. For instance, the Wald entropy and Bekenstein entropy do not align in Gauss-Bonnet gravity.}
\begin{equation}
S=\int _0^{r_h} \frac{1}{T}dM=\pi  r_h^2+4 \pi \log (r_h) .
\end{equation}
Additionally, we obtain the thermodynamic volume,
\begin{equation}
V=\left( \frac{\partial M}{\partial P}\right) _{S,Q,\alpha}=\frac{4}{3} \pi  r_h^3.
\end{equation}
We observe that the entropy includes a logarithmic correction term, while the thermodynamic volume remains equal to the geometric volume. The corresponding Smarr relation,
\begin{equation}
    M=2TS+\Phi Q -2PV +2\alpha \mathcal{A}
\end{equation}
can be derived from it through a scaling (dimensional) argument \cite{Kastor:2009wy, Wei:2020poh}. When $P$ is considered constant (i.e., if the cosmological constant is not allowed to vary), \eqref{firstlaw} reduces to the standard first law in the 'non-extended' phase space. However, in this scenario, the Smarr relation remains unchanged and no longer follows directly from the first law through the scaling argument.


\section{Phase Transition of the Black Hole}
\label{sectwo}
With the definitions of the thermodynamic quantities and the first law in place, it is now straightforward to analyze the phase structure of the black hole. We will explore the phase transition properties by examining the $P-v$ diagram, the behavior of Gibbs free energy, and the specific heat of the system. The equation of state can be readily obtained as,
\begin{equation}
P=\frac{Q^2}{8 \pi  r_h^4}+\frac{\alpha }{8 \pi  r_h^4}+\frac{\alpha  T}{r_h^3}-\frac{1}{8 \pi  r_h^2}+\frac{T}{2 r_h}.
\end{equation}
When $\alpha =0$, this equation reduces to the equation of state of the charged AdS black hole \cite{Kubiznak2012}. The functional behavior of this equation of state is similar to that of the van der Waals case. To ensure dimensional correctness, we rescale it as follows,
\begin{equation}
P\rightarrow \frac{\hbar c}{l_P^2}, \quad T\rightarrow \frac{\hbar c}{k_B}T.
\end{equation}
Here, $\hbar$, $c$, $k_B$, and $l_P$ have usual meanings. Comparing these with the van der Waals equation of state, the relationship between the specific volume $v$ and the horizon radius is $v=2l_P r_h$. As demonstrated below, the $4D$ Gauss-Bonnet AdS black hole exhibits a phase structure similar to that of the van der Waals system. The critical point associated with this phase structure is characterized by,
\begin{equation}
\left( \frac{\partial P}{\partial v}\right)_T =\left( \frac{\partial ^2P}{\partial v^2}\right) _T=0.
\end{equation}
The critical temperature, critical pressure, and critical volume are obtained from these conditions,
\begin{equation}
T_c=\frac{\left(8 \alpha +3 Q^2-\rho \right) \sqrt{6 \alpha +3 Q^2+\rho }}{48 \pi  \alpha ^2};
\end{equation}
\begin{equation}
P_c=\frac{9 \alpha +6 Q^2+\rho }{24 \pi  \left(6 \alpha +3 Q^2+\rho \right)^2};
\end{equation}
\begin{equation}
V_c=\frac{4}{3} \pi  \left(6 \alpha +3 Q^2+\rho \right)^{3/2};
\end{equation}
where $\rho =\sqrt{48 \alpha ^2+9 Q^4+48 \alpha  Q^2}$. 

\begin{figure*}[t]
\centering
\subfigure[ref2][]{\includegraphics[width=0.47\textwidth]{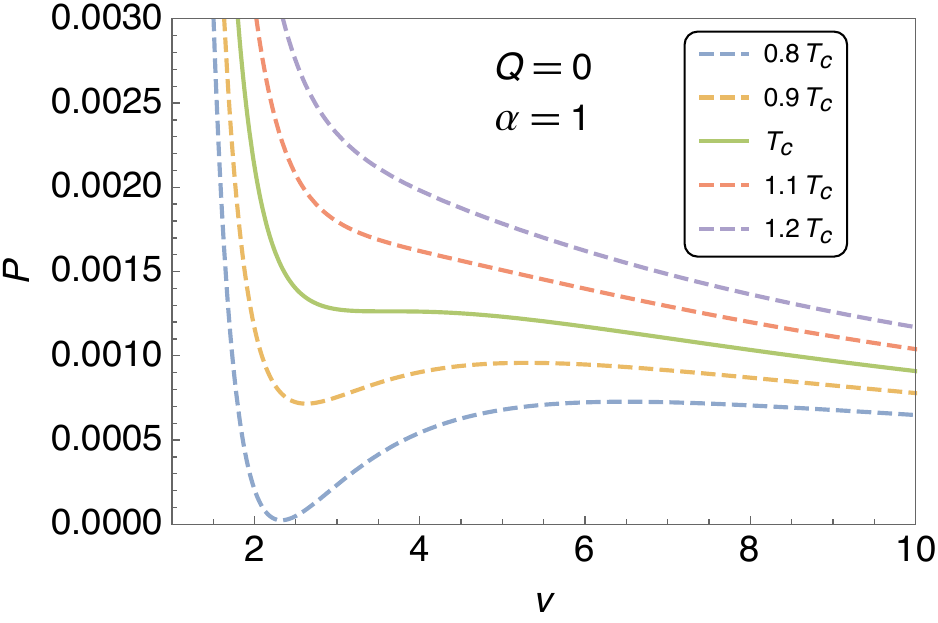}\label{PVuncharged}}
\qquad
\subfigure[ref1][]{\includegraphics[width=0.47\textwidth]{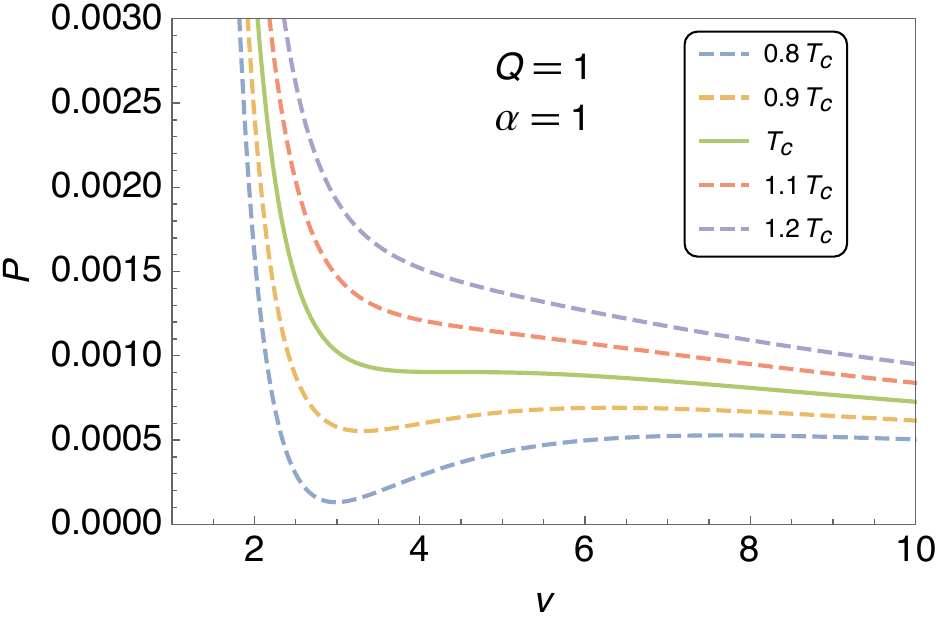}\label{PVcharged}}
\caption{$P-V$ isotherms for neutral (left) and charged (right) $4D$ Gauss-Bonnet AdS black holes. }
\label{PV}
\end{figure*}

 In all these expressions written so far with $Q$, the neutral case is obtained by setting $Q=0$. The $P-v$ diagram is shown in figure \ref{PV}, which exhibits critical behavior. The plots display qualitatively different behaviors for temperatures above and below critical values. The isotherms above the critical temperature $T_c$ are injective, with only one dominant black hole solution in the phase diagram for large temperatures. Below the critical temperature $T_c$, the isotherms feature three distinct regions. The regions with negative slope represent stable phases, while those with positive slope denote unstable phases. A first-order phase transition occurs between the stable phases, namely small black hole (SBH) and large black hole (LBH). The unstable region disappears for temperatures above the critical value. Notably, the critical behavior is observed for both neutral and charged Gauss-Bonnet black holes. This finding is intriguing because, for four-dimensional RN AdS black holes, charge is the key property for displaying van der Waals-like phase transitions. The presence of a van der Waals-like phase transition in the neutral case suggests that the Gauss-Bonnet parameter $\alpha$ assumes a role similar to that of charge, as observed in five-dimensional Gauss-Bonnet AdS black holes \cite{Cai:2013qga}.

\begin{figure*}[t]
\centering
\subfigure[ref2][]{\includegraphics[width=0.47\textwidth]{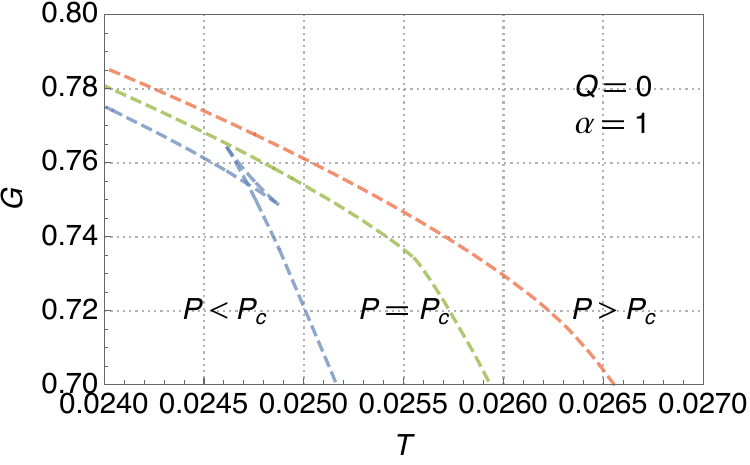}\label{GTuncharged}}
\qquad
\subfigure[ref1][]{\includegraphics[width=0.47\textwidth]{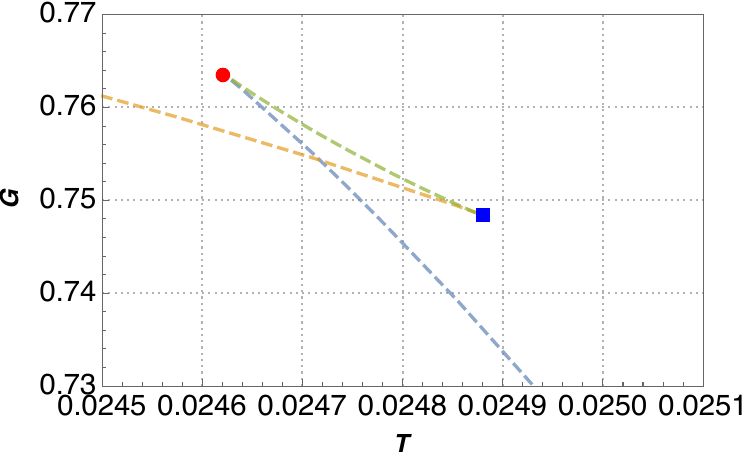}\label{GTcharged}}
\caption{The Gibbs free energy $G$ vs Hawking temperature $T$ plot for different pressure values corresponding to above and below the critical point is depicted for neutral  $4D$ Gauss-Bonnet AdS black hole (left). In the right panel we show the enlarged view of the swallow tail behavior. (Charged case also exhibits similar behavior).}
\label{GT}
\end{figure*}

The phase transition can be better understood by investigating the behavior of Gibbs free energy. The Gibbs free energy $G$ of the black hole is calculated using the standard expression $G=M-TS$, which yields \footnote{In extended black hole thermodynamics, the black hole mass $M$ is analogous to the thermodynamic potential, enthalpy, rather than the energy as in conventional black hole thermodynamics.},
\begin{eqnarray}
G=\frac{1}{6} \left[\frac{3 \left(r_h^2+4 \alpha  \log (r_h)\right) \left(\alpha -8 \pi  P r_h^4+Q^2-r_h^2\right)}{2 \left(r_h^3+2 \alpha  r_h\right)}  +8 \pi  P r_h^3+\frac{3 Q^2}{r_h}+\frac{3 \alpha }{r_h}+3 r_h\right].
\end{eqnarray}
Here, $r_h$ is considered a function of $(P,T)$ from the equation of state. In the free energy plot shown in Fig. \ref{GT}, for temperatures $T<T_c$, three branches exist. The first branch starts from the left side and ends at the red dot, representing the small black hole phase (SBH). The second branch starts from this red dot and ends at the blue square, corresponding to the intermediate black hole phase (IBH). The third branch starts from this blue square and continues downwards as the temperature increases, representing the large black hole phase (LBH). At the point where the SBH and LBH branches meet, a first-order SBH-LBH phase transition of the black hole occurs. In other words, the presence of this characteristic swallowtail behavior for both neutral and charged Gauss-Bonnet black holes is the signature of the first-order phase transition of the system. This first-order phase transition point depends on the value of $P$, implying that it is possible to represent the phase structure with a coexistence curve in the $P-T$ plane. In contrast to the five-dimensional uncharged case \cite{Mo:2016sel}, obtaining an analytical expression for coexistence curve for the $4D$ Gauss-Bonnet black hole is not feasible. The coexistence curve separates the SBH and LBH phases below the critical point. The IBH branch is never dominant and hence does not play a role in the phase diagram. As we will see below, the heat capacity of this phase of the black hole is always negative, making it thermodynamically unstable. For $P=P_c$, the IBH branch disappears, and hence the SBH and LBH branches merge. This corresponds to the termination point of the $P-T$ coexistence curve, and the phase transition at this point is second order. Finally, when $P>P_c$, there is no phase transition.

\begin{figure*}[t]
\centering
\subfigure[ref2][]{\includegraphics[width=0.47\textwidth]{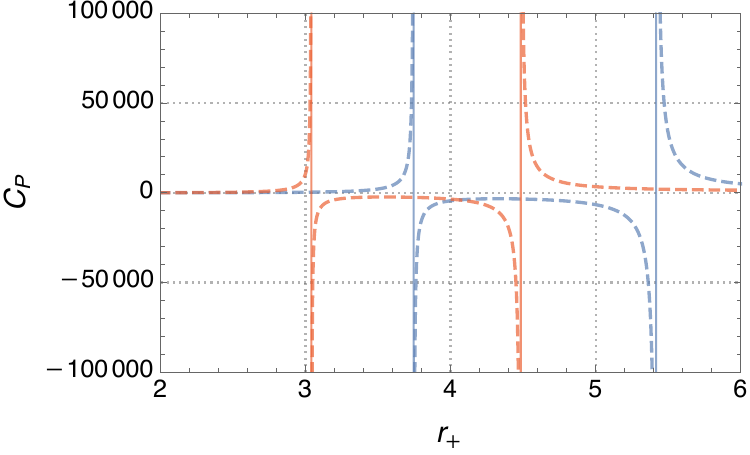}\label{CPbelow}}
\qquad
\subfigure[ref1][]{\includegraphics[width=0.47\textwidth]{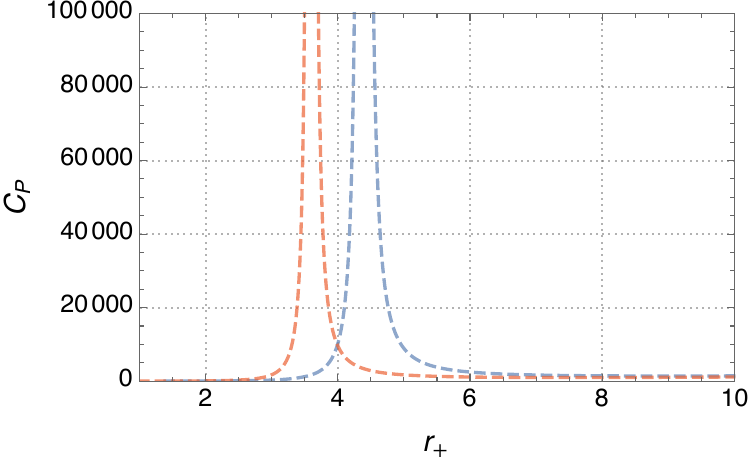}\label{CPcritical}}

\subfigure[ref1][]{\includegraphics[width=0.47\textwidth]{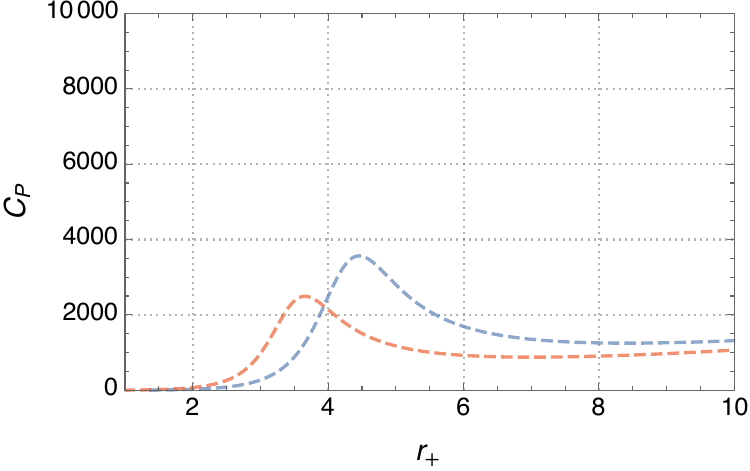}\label{CPabove}}
\caption{The specific heat behavior of the $4D$ Gauss-Bonnet AdS black hole is illustrated. The blue (solid) lines represent the charged black hole, while the red (dashed) curves represent the neutral black hole. Fig. \ref{CPbelow} corresponds to $P<P_c$, Fig. \ref{CPcritical} corresponds to $P=P_c$, and Fig. \ref{CPabove} corresponds to $P>P_c$. In the charged case, $Q=1$, and in both neutral and charged cases, $\alpha=1$.}
\label{CP}
\end{figure*}

Another crucial thermodynamic quantity to explore in the context of phase transitions and thermal stability is the specific heat. A positive heat capacity corresponds to a stable system, whereas, a negative heat capacity indicates the instability of the system under small perturbations. We examine the specific heat at constant pressure,
\begin{equation}
C_P=T\left( \frac{\partial S}{\partial T}\right)_P. 
\end{equation}
For the $4D$ Gauss-Bonnet AdS black hole, we obtain
\begin{equation}
C_P=\frac{2 \pi  \left(2 \alpha +r_h^2\right)^2 \left(-\alpha +8 \pi  P r_h^4-Q^2+r_h^2\right)}{2 \alpha ^2+8 \pi  P r_h^6+r_h^4 (48 \pi  \alpha  P-1)+Q^2 \left(2 \alpha +3 r_h^2\right)+5 \alpha  r_h^2}.
\end{equation}
The behavior of $C_P$ is depicted in Fig. \ref{CP}.

We have simultaneously considered charged and neutral black holes since there is no difference in the characteristic behavior of the specific heat between both cases. However, quantitative changes are apparent from the plots. For $P<P_c$, two diverging points exist, which separate three regions. The phases with positive specific heat in the lower radius and higher radius regions are stable. The intermediate phase with a negative $C_P$ value is an unstable phase. Therefore, for pressures lower than the critical value, three phases are possible: small black hole (SBH), intermediate black hole (IBH), and large black hole (LBH). The phase transition occurs between the stable phases, SBH and LBH. When $P=P_c$, the two divergent points merge to form a single divergence, removing the unstable region. For $P>P_c$, the heat capacity is always positive, and there are no divergences. This implies that the black hole is stable, and there is no phase transition. Indeed, these results bear resemblance to those observed in RN AdS black holes in four dimensions and Gauss-Bonnet theory in higher dimensions.

We can better understand the stability of black holes in the free energy plot by recasting the expression for heat capacity as follows. Using the thermodynamic identity $S=-\partial G/\partial T$,
\begin{equation}
C_P=-T\left( \frac{\partial ^2 G}{\partial T^2}\right)_P
\end{equation}
In the $G-T$ plot, a branch is thermodynamically unstable if it is concave upwards and stable if it is concave downwards. From Fig. \ref{GT}, for the $P<P_c$ case, it is clear that the two intersecting branches (first and third) correspond to stable phases (SBH and LBH). The remaining branch corresponds to the unstable branch (IBH).

An important set of quantities that characterize the critical behavior are the critical exponents. Using the method prescribed in reference \cite{Kubiznak2012}, we obtain the critical exponents as $\alpha =0$, $\beta =1/2$, $\gamma =1$, and $\delta =3$, which are universal.


\subsection{Order parameter of the phase transition}
To further understand the van der Waals-like phase structure associated with Gauss-Bonnet spacetime, we introduce a new order parameter. From the phase transition properties of the Gauss-Bonnet black hole, it is evident that the Gauss-Bonnet parameter $\alpha$ plays an analogous role to black hole charge or spin in determining the black hole phase structure. To emphasize this, we focus on the neutral Gauss-Bonnet black hole in this section. First, we analyze the behavior of the conjugate variable $\mathcal{A}$ against the Hawking temperature. The explicit expression for $\mathcal{A}$ is given by \cite{Wei:2020poh}.
\begin{equation}
 \mathcal{A}=\left(\frac{\partial M}{\partial \alpha}\right)_{S, Q, P}=\frac{\alpha +2 \ln
   \left(\frac{r_h}{\sqrt{\alpha
   }}\right) \left(\alpha -8 \pi  P
   r_h^4-r_h^2+Q^2\right)+8 \pi  P
   r_h^4+2 r_h^2-Q^2}{2 \left(2 \alpha
   r_h+r_h^3\right)}.
   \label{Aeqn}
\end{equation}

\begin{figure}[t]
\centering
\subfigure[ref3][]{\includegraphics[width=0.47\textwidth]{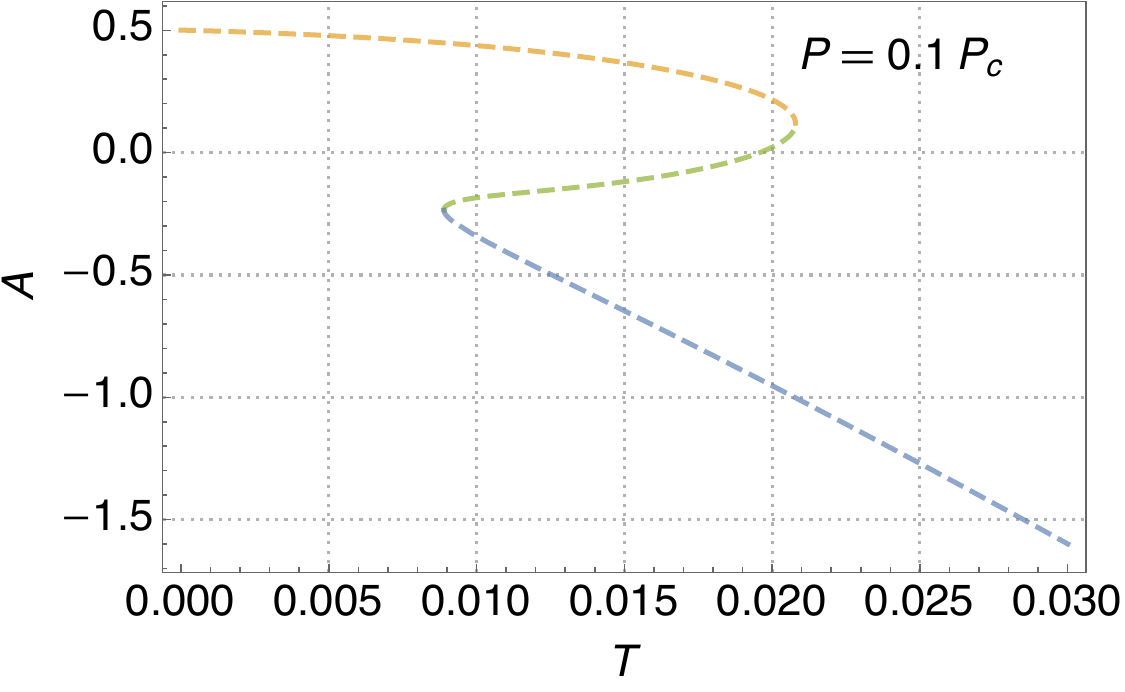}\label{ATplot1}}
\qquad
\subfigure[ref4][]{\includegraphics[width=0.47\textwidth]{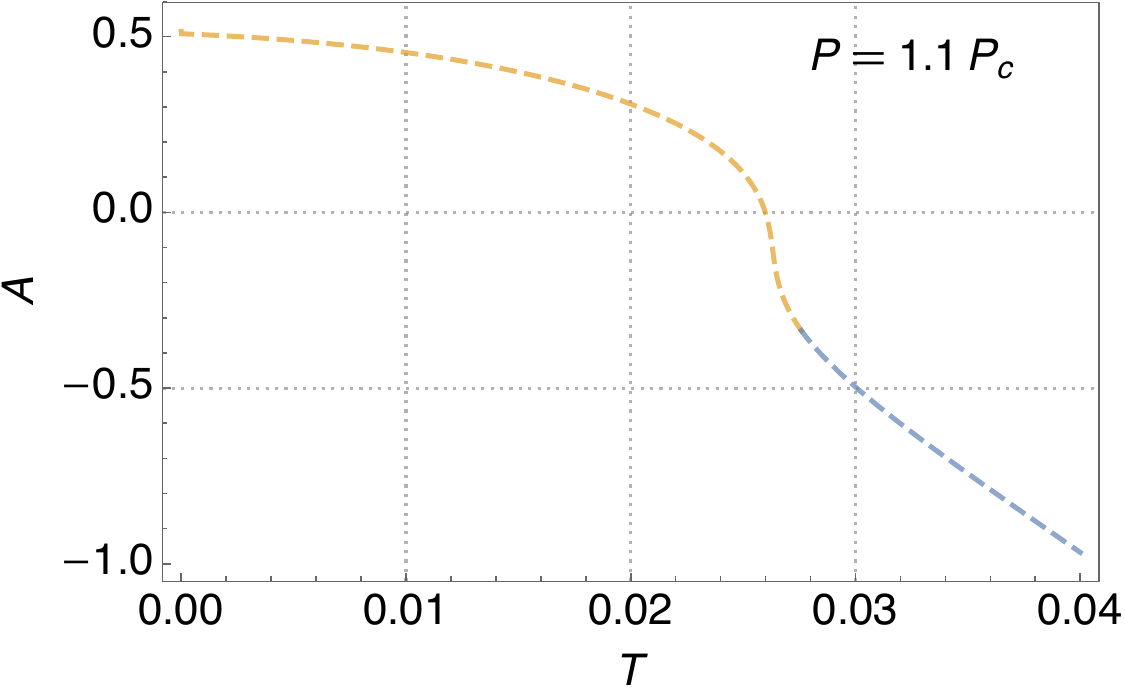}\label{ATplot2}}
\caption{The behavior of the conjugate variable $\mathcal{A}$ with respect to the Hawking temperature $T$ is depicted for pressure values corresponding to below (left panel) and above (right panel) the critical point of the black hole phase transition.}
\label{APplots}
\end{figure}

Using \eqref{Hawking}, we express $r_h$ as a function of $T$. In the parameter domain where the black hole undergoes a phase transition, $r_h(T)$ is multivalued, corresponding to multiple phases of the black hole. Substituting $r_{h}(T)$ into \eqref{Aeqn}, we express $\mathcal{A}$ in terms of $T$. In Fig. \ref{APplots}, $\mathcal{A}$ is plotted against $T$ for two different values of $P$. For $P<P_c$, three black hole solutions (Small BH, Intermediate BH, and Large BH) coexist for $T_1 < T < T_2$. Moreover, the first-order phase transition between Small BH and Large BH occurs at $T_p$. When $T_1 < T < T_2$, $\mathcal{A}$ possesses three branches, which exactly match Small BH, Intermediate BH, and Large BH, respectively. The right column displays the case of $P>P_c$, where there is only one black hole solution and no phase transition. These observations suggest that $\mathcal{A}$ can be used to probe the phase structure of Gauss-Bonnet black holes.

\begin{figure}[t]
    \centering
    \includegraphics[width=0.47\textwidth]{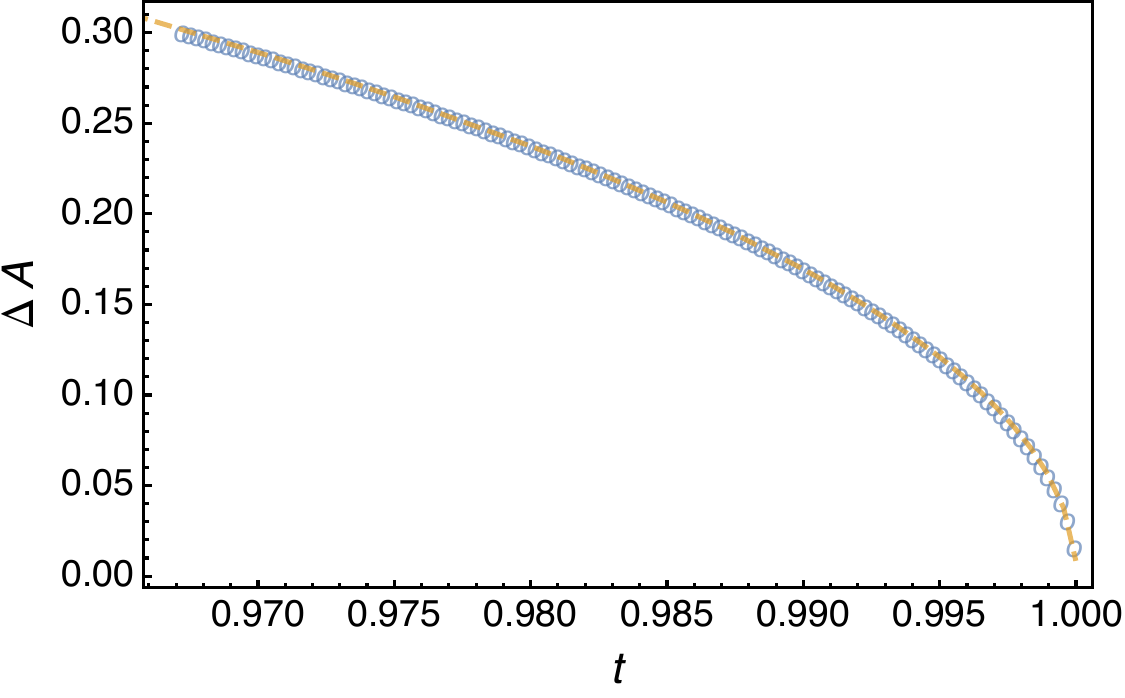}
    \caption{The plot of $\Delta \mathcal{A}$ against $t$ near the critical point. }
    \label{fig:orderparameter}
\end{figure}

Indeed, the phase transition can be characterized by the discontinuous change in the conjugate variable $\mathcal{A}$, expressed as $\Delta \mathcal{A} = \mathcal{A} _S - \mathcal{A} _L$, where $\mathcal{A} _S$ and $\mathcal{A} _L$ represent the $\mathcal{A}$ values of Small BH and Large BH evaluated at $T = T_P$, respectively. Notably, for the second-order phase transition at the critical point, $\mathcal{A} _S = \mathcal{A} _L = \mathcal{A} _c$, resulting in $\Delta \mathcal{A} = 0$. In Figure \ref{fig:orderparameter}, we illustrate $\Delta \mathcal{A} /\mathcal{A} _c$ as a function of $t \equiv T_p /T_c$, where $t = 1$ at the critical point. It is evident that during the first-order phase transition from Small BH to Large BH, the conjugate variable $\mathcal{A}$ jumps from $\mathcal{A} _S$ to $\mathcal{A} _L$ with a nonzero $\Delta \mathcal{A} $. Consequently, $\Delta \mathcal{A} $ can be treated as an order parameter. To investigate the critical behavior of $\Delta \mathcal{A} $, we expand $\Delta \mathcal{A} $ in terms of $t$ near the critical point and obtain,
\begin{equation}
   \frac{\Delta \mathcal{A} }{\mathcal{A} _c } \propto \sqrt{1-t}
\end{equation}
This calculation yields a critical exponent of $\Delta \mathcal{A}$ as 1/2. Remarkably, our findings indicate that the critical exponent of $\Delta \mathcal{A}$ mirrors that of the order parameter in the van der Waals fluid as predicted by mean field theory.

\section{Joule Thomson expansion of the black hole}
\label{secthree}
An intriguing aspect of a van der Waals (vdW) system is the occurrence of heating or cooling during the throttling process. Given the analogy between the phase structures of $4D$ Gauss-Bonnet black hole systems and vdW systems, it is worthwhile to explore the Joule-Thomson expansion of the black hole.

\subsection{Joule-Thomson effect}

The Joule-Thomson effect involves the irreversible adiabatic expansion of a gas as it passes through a porous plug. In this process, a non-ideal gas undergoes continuous throttling, resulting in a temperature change in the final state. When gas from a higher-pressure side with pressure $P_i$ and temperature $T_i$ expands through a porous plug, it traverses dissipative non-equilibrium states due to friction between the gas and the plug. Usual thermodynamic coordinates are inadequate for describing these non-equilibrium states. However, it has been observed that the enthalpy, defined as the sum of internal energy and the product of pressure and volume, remains constant in the final state \citep{Zemansky2011}. Thus, the enthalpy function $H=U+PV$ is introduced as a state function, which remains unchanged in the final states
\begin{equation}
H_i= H_f
\label{H}
\end{equation}
It is not accurate to assert that enthalpy remains constant during this process, as enthalpy is not defined when the gas undergoes non-equilibrium states. The phase diagram consists of a set of discrete points, with the initial point $(P_i, T_i)$ and all subsequent points $(P_f, T_f)$ representing equilibrium states of a gas with the same molar enthalpy $(h)$ at both the initial and final equilibrium states. These discrete points, corresponding to the same molar enthalpy, lie on a smooth curve known as an isenthalpic curve. In summary, an isenthalpic curve represents the locus of all points with the same molar enthalpy, encompassing both initial and final equilibrium states. Multiple isenthalpic curves can be obtained for different values of enthalpy.

The slope of an isenthalpic curve on the $T-P$ plane is referred to as the \emph{Joule-Thomson coefficient} $\mu _J$, defined as
\begin{equation}
\mu _J=\left( \frac{\partial T}{\partial P}\right) _H.
\end{equation}
The Joule-Thomson coefficient is zero at the maxima of the isenthalpic curve, and the locus of such points is called the inversion curve. Inside the inversion curve, where the gradient of isenthalps ($\mu _J$) is positive, is termed the region of cooling, while outside, where $\mu _J$ is negative, is known as the region of heating. The differential of molar enthalpy is given by
\begin{equation}
dh=Tds+vdP.
\label{dh}
\end{equation}
Recalling the second $TdS$ equation in classical thermodynamics
\begin{equation}
TdS=C_PdT-T\left( \frac{\partial v}{\partial T}\right) _PdP.
\label{tds}
\end{equation}
Using Equation (\ref{tds}) in Equation (\ref{dh}) yields
\begin{equation}
dT=\frac{1}{C_P}\left[ T\left( \frac{\partial v}{\partial T}\right) _P-v\right]dP+\frac{1}{C_P}dh.
\end{equation}
This gives
\begin{equation}
\mu _J=\left( \frac{\partial T}{\partial P}\right) _H=\frac{1}{C_P}\left[ T\left( \frac{\partial v}{\partial T}\right) _P-v\right].
\end{equation}
The condition $\mu _J=0$ defines the inversion temperature.


\subsection{Joule Thomson expansion of Gauss-Bonnet AdS blackhole}

In the extended phase space, the black hole mass is equivalent to enthalpy \cite{Kubiznak:2016qmn, Dolan:2011xt}. Hence, the isenthalpic curves, representing the locus of all points corresponding to the initial and final equilibrium states of the same enthalpy, are constant mass curves. The slope of an isenthalpic curve is the Joule-Thomson coefficient, which now reads,
\begin{equation}
\mu _J =\left( \frac{\partial T}{\partial P}\right) _M=\frac{1}{C_P}\left[ T\left( \frac{\partial V}{\partial T}\right) _P-V\right].
\end{equation}
As $\mu _J =0$ defines the inversion temperature, we have
\begin{equation}
T_i=V\left( \frac{\partial T}{\partial V}\right) _P.
\label{eqJT}
\end{equation}
For the Gauss-Bonnet AdS black hole, we obtain,
\begin{equation}
T_i=\frac{2 \alpha ^2+8 \pi  P \left(r^6+6 \alpha  r^4\right)+Q^2 \left(2 \alpha +3 r^2\right)-r^4+5 \alpha  r^2}{12 \pi  r \left(2 \alpha +r^2\right)^2}
\label{eqstate1}
\end{equation}
From Equation (\ref{Hawking}), we have
\begin{equation}
T_i=-\frac{\alpha -8 \pi  P r^4+Q^2-r^2}{4 \pi  r^3+8 \pi  \alpha  r}.
\label{eqstate2}
\end{equation}
\begin{figure*}[t]
\centering
\subfigure[ref2][]{\includegraphics[width=0.47\textwidth]{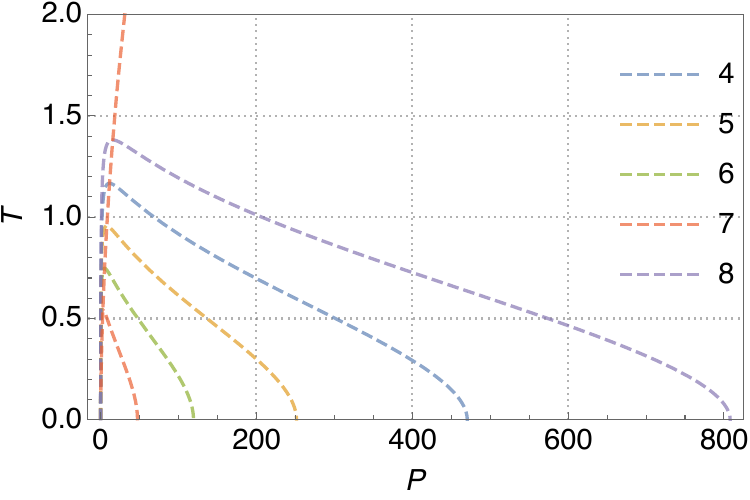}\label{GBJTuncharged}}
\qquad
\subfigure[ref1][]{\includegraphics[width=0.47\textwidth]{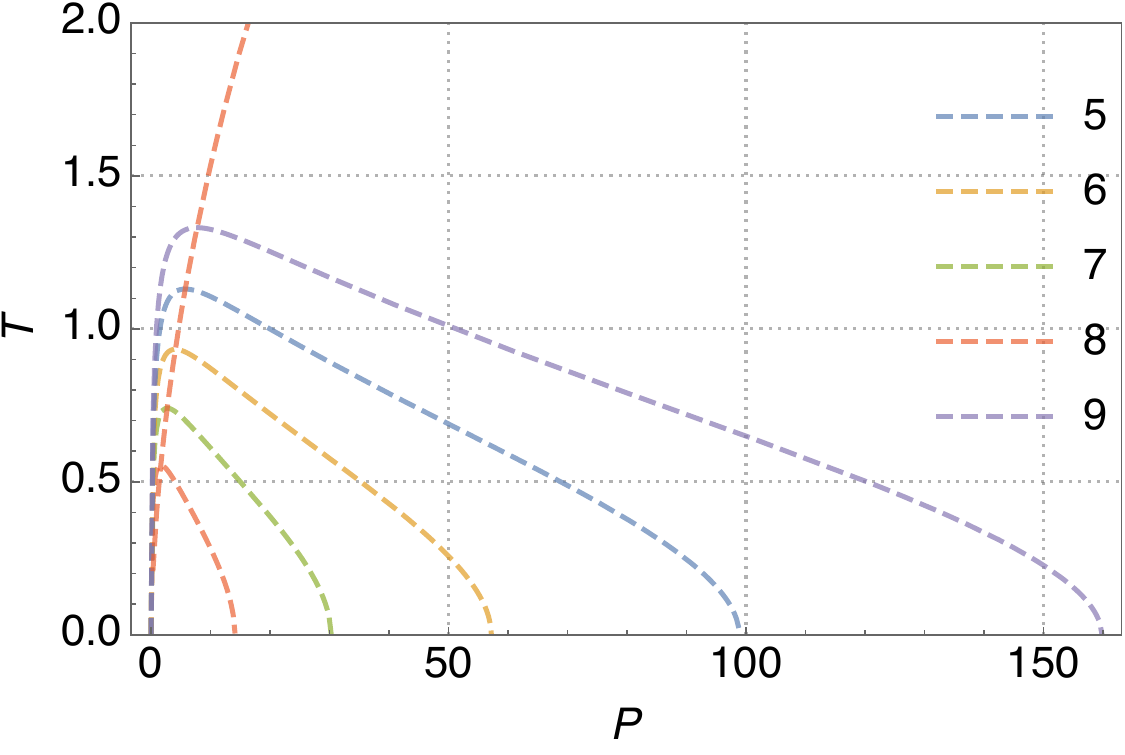}\label{GBJTcharged}}
\caption{Inversion curve and isenthalpic curves for neutral (left) and charged (right) $4D$ Gauss-Bonnet AdS black holes. The red monotonously increasing line represents the inversion curve. The curves with changing slopes, exhibiting maxima on the inversion curve, depict the isenthalps for different values of $M$.}
\label{GBJT}
\end{figure*}

We solve equations (\ref{eqstate1}) and (\ref{eqstate2}) for $r_h$  and obtain
\begin{equation} \label{eq30}
   r_h= \frac{1}{2 \sqrt{3 \pi }}\sqrt{\frac{\Gamma ^{2/3}-\Gamma ^{1/3}+6 \pi  P \left(\alpha +3 Q^2\right)+1}{\Gamma ^{1/3} P}}
\end{equation}
where
\begin{equation}
\begin{split}
    \Gamma &=  432 \pi ^2 \alpha  P^2 \left(\alpha +Q^2\right)-9 \pi  P \left(\alpha +3 Q^2\right)-1\\
    &+\sqrt{\left(-432 \pi ^2 \alpha  P^2 \left(\alpha +Q^2\right)+9 \pi  P \left(\alpha     +3 Q^2\right)  +1\right)^2- \left(6 \pi  P \left(\alpha +3 Q^2\right)+1\right)^3}.
    \end{split}
\end{equation}
By substituting \eqref{eq30} into equation \eqref{eqstate1}, we obtain the expression for the inversion temperature,
\begin{equation}
\begin{split}
  T_i=&  \frac{1}{6 \sqrt{3 \pi } \Gamma ^{1/3} \left(\Gamma ^{2/3}-\Gamma ^{1/3}+6 \pi  P_i \left(4 \alpha  \Gamma ^{1/3}+\alpha +3 Q^2\right)+1\right)^2 \sqrt{\frac{\Gamma ^{2/3}-\Gamma ^{1/3}+6 \pi  P_i \left(\alpha +3 Q^2\right)+1}{\Gamma ^{1/3} P_i}}}\\
 \times &   \Biggl[ 2 \Gamma ^2+9 \Gamma ^{5/3} (16 \pi  \alpha  P_i-1)+\Gamma  \left(144 \pi  P_i \left(\alpha +18 \pi  \alpha ^2 P_i+Q^2 (42 \pi  \alpha  P_i-3)\right)-23\right) \\
  & +18 \Gamma ^{2/3} \left(4 \pi  P_i \left(3 Q^2-\alpha \right)+1\right) \left(6 \pi  P_i \left(\alpha +3 Q^2\right)+1\right)+18 \Gamma ^{4/3} \left(4 \pi  P_i \left(3 Q^2-\alpha \right)+1\right) \\
  & +9 \Gamma ^{1/3} (16 \pi  \alpha  P_i-1) \left(6 \pi  P_i \left(\alpha +3 Q^2\right)+1\right)^2+2 \left(6 \pi  P_i \left(\alpha +3 Q^2\right)+1\right)^3 \Biggr]
   \end{split}
\end{equation}

We plot this in Figure \ref{GBJT}. Both neutral and charged Gauss-Bonnet black holes exhibit similar behavior, reaffirming our previous observation that the Gauss-Bonnet coupling parameter plays a role analogous to the charge or spin of a black hole in determining its phase structure. In other words, the presence of an electric charge $Q$ is not necessary to observe the Joule Thomson expansion phenomenon in Gauss-Bonnet spacetime. Instead, the Joule Thomson expansion is guided by the Gauss-Bonnet coupling constant $\alpha$. The isenthalps are also shown in the same set of diagrams, known as crossing diagrams. The region to the left of the inversion curve corresponds to the cooling region, while to the right is the heating region during the throttling process.


\section{Conclusions}
\label{secfour}
In this article, we have examined the thermodynamics and phase transition properties of asymptotically AdS black holes within the Einstein-Gauss-Bonnet gravity, with a particular focus on the Joule-Thomson expansion. While both charged and neutral cases exhibit similar behavior, the neutral case holds particular interest.  We observed that several thermodynamic characteristics of $4D$ Gauss-Bonnet AdS black holes resemble those of charged and spinning black holes in AdS spacetime, a similarity also observed in higher-dimensional Gauss-Bonnet black holes. This underscores the unique influence of the Gauss-Bonnet parameter on the thermodynamic description. It is well-known that Schwarzschild AdS black holes exhibit Hawking-Page transitions, while charged/spinning AdS black holes demonstrate van der Waals-like phase transitions. Given that neutral $4D$ Gauss-Bonnet AdS black holes display van der Waals-like phase transitions, we can infer that Gauss-Bonnet parameter induces effects similar to those of black hole charge or spin in guiding the black hole phase structure.

We first explored the extended thermodynamics of black holes, treating the cosmological constant as pressure. We studied the phase transition of the black hole  through the $P-V$ criticality, Gibbs free energy, and specific heat behaviors. The black hole was observed to undergo a first-order phase transition between a small black hole phase and a large black hole phase. This first-order transition line terminates at the critical point, where the phase transition becomes second order. We carefully analyse how the Gauss-Bonnet parameter influences the phase structure of the Gauss-Bonnet AdS black hole, particularly by examining the behavior of the conjugate variable to the Gauss-Bonnet parameter, denoted as $\mathcal{A}$. It was found that the behavior of $\mathcal{A}$ as a function of temperature below the critical point ($P<P_c$) demonstrates three branches, corresponding to three coexisting black hole phases. Conversely, when the pressure exceeds the critical value ($P>P_c$), $\mathcal{A}$ becomes a single-valued function of temperature, coinciding with one black hole phase. At the first-order phase transition, the discontinuity in $\Delta \mathcal{A}$ can serve as an order parameter to characterize the black hole phase transition. Notably, $\Delta \mathcal{A}$ exhibits a critical exponent of $1/2$ at the critical point.

In the second part of the article, we analyse the Joule Thomson expansion of the $4D$ Gauss-Bonnet black hole. Since the mass of the black hole is identified as the enthalpy of the system in the extended thermodynamics, it remains constant throughout the Joule Thomson expansion. We derived an analytical expression for the inversion temperature of the $4D$ Gauss-Bonnet AdS black hole, unveiling insights into its throttling process, where expansion can induce either heating or cooling effects. Through the illustration of cooling and heating regions across various charge ($Q$) and mass ($M$) values, we observed the presence of Joule Thomson expansion in both charged and neutral Gauss-Bonnet black holes. This observation highlights the analogous role of the Gauss-Bonnet parameter to black hole charge or spin in governing the black hole phase structure in Gauss-Bonnet spacetime. Given that $\mathcal{A}$ serves as an order parameter, an intriguing aspect for further exploration lies in investigating the microstructure and continuous phase transition using the Landau theory \cite{NaveenaKumara:2020lgq, Guo:2019oad}. Additionally, examining the holographic dual of extended black hole thermodynamics in this context for the $4D$ Gauss-Bonnet AdS black hole \cite{Ahmed:2023snm} could offer valuable insights from a holographic standpoint.

Another avenue for exploration in the context of 4D Einstein-Gauss-Bonnet black holes is the study of the correlation between thermodynamic (in)stability and dynamical (in)stability. Conventionally, in the study of gravitational perturbations, larger multipole numbers $\ell$ are expected to lead to a higher barrier of the effective potential, thus stabilizing the system. However, in theories with higher curvature corrections, larger $\ell$ can correspond to deeper negative gaps near the event horizon, leading to a new instability known as eikonal (gravitational) instability \cite{Konoplya:2017ymp, Cuyubamba:2016cug}. The eikonal instability of 4D Einstein-Gauss-Bonnet black holes has been studied in Ref. \cite{Konoplya:2020juj}, which identified parametric regions of eikonal instability for various values of coupling and cosmological constants. Exploring the connection between the thermal and dynamical stability of black holes is an intriguing aspect (see, for example, \cite{Hollands:2012sf}), and we believe that extending the work in this direction, particularly in the context of extended thermodynamics, would be highly valuable.

\section*{Acknowledgements}
Authors K.H., N.K.A. and A.R.C.L. would like to thank U.G.C. Govt. of India for financial assistance under UGC-NET-SRF scheme.


\bibliographystyle{elsarticle-num} 
\bibliography{BibTex.bib}

\end{document}